\documentclass{aa}
\usepackage{txfonts}
\usepackage{graphicx}

\begin{document}

\title{The elusive tidal tails of the Milky Way globular cluster NGC\,7099\thanks{DECam 
photometric data are only available in electronic form
at the CDS via anonymous ftp to cdsarc.u-strasbg.fr (130.79.128.5).}}

\author{Andr\'es E. Piatti\inst{1,2}\thanks{\email{andres.piatti@unc.edu.ar}}, Julio A. Carballo-Bello\inst{3}, Marcelo D. Mora\inst{4}, Carolina Cenzano\inst{4}, Camila Navarrete\inst{5,6},\\
and M\'arcio Catelan\inst{4,6}
}

\institute{Instituto Interdisciplinario de Ciencias B\'asicas (ICB), CONICET-UNCUYO, Padre J. Contreras 1300, M5502JMA, Mendoza, Argentina
\and Consejo Nacional de Investigaciones Cient\'{\i}ficas y T\'ecnicas (CONICET), Godoy Cruz 2290, C1425FQB,  Buenos Aires, Argentina
\and Instituto de Alta Investigaci\'on, Universidad de Tarapac\'a, Casilla 7D, Arica, Chile
\and
Instituto de Astrof\'isica, Facultad de F\'isica, Pontificia Universidad Cat\'olica de Chile, Av. Vicu\~na Mackenna 4860, 782-0436 Macul, Santiago, Chile
\and
European Southern Observatory, Alonso de C\'ordova 3107, Casilla 19001, Santiago, Chile
\and
Millenium Institute of Astrophysics, Santiago, Chile\\
}

\date{Received / Accepted}

\abstract{We present results on the extra-tidal features of the Milky Way globular cluster NGC\,7099,
using deep $gr$ photometry obtained with the Dark Energy Camera (DECam). We reached
nearly 6 mag below the cluster Main Sequence (MS) turnoff, so that we dealt with the most
suitable candidates to trace any stellar structure located beyond the cluster tidal radius.
From star-by-star reddening corrected color-magnitude diagrams (CMDs) we defined four adjacent
strips along the MS, for which we built the respective stellar density maps, once the contamination by field stars was properly removed. The resulting field star cleaned stellar density maps show 
a short tidal tail and some scattered debris.
Such extra-tidal features are hardly detected when much shallower {\it Gaia} DR2 data sets 
are used and the same CMD field star cleaning procedure is applied. Indeed, by using 
2.5 magnitudes below the cluster MS turnoff as the faintest limit  ($G <$ 20.5 mag), cluster 
members turned out to be distributed within the cluster's tidal radius, and some hints for field star 
density variations are found  across a circle of radius 3.5$\degr$ centered on the cluster 
and with similar CMD features as cluster stars. The proper motion distribution of these stars is
distinguishable from that of the cluster, with some superposition, which resembles that of stars
located beyond 3.5$\degr$ from the cluster center.

}
 

\titlerunning{The tidal tails in NGC\,7099}

\authorrunning{A.E. Piatti et al.}

\maketitle

\markboth{A.E. Piatti et al.: }{The tidal tails in NGC\,7099}

\section{Introduction}

Tidal streams are witnesses of accreted dwarf galaxies disrupted by the Milky Way. They have been
unveiled from wide-sky photometric and spectroscopic surveys, including Sloan Digital Sky Survey
(SDSS)\footnote{https://www.sdss.org}, Two Micron All-Sky Survey (2MASS)\footnote{https://irsa.ipac.caltech.edu/Missions/2mass.html}, Panoramic Survey Telescope and Rapid Response System (Pan-STARRS)\footnote{https://panstarrs.stsci.edu}, Dark Energy Survey
(DES)\footnote{https://www.darkenergysurvey.org} and 
{\it Gaia} mission\footnote{https://gea.esac.esa.int/archive/}. The most spectacular example of these stellar 
substructures in the Milky Way halo is the one generated by the assimilation of the Sagittarius dwarf
galaxy, which is moving around the Milky Way in an almost polar orbit \citep{ibataetal1994,majewskietal2003,belokurovetal2006b,koposovetal2012}. Additional 
well-known streams in the Milky Way include the Monoceros ring, which may represent an on-plane 
accretion of a minor satellite, and the Sausage-Enceladus structure, which has been recently 
discovered in the {\it Gaia} DR2 \citep{belokurovetal2018,helmietal2018} and seems to be associated 
with the impact of an Small Magellanic Cloud-like mass galaxy on the Milky Way. 
There are also minor streams reported by \citet{mateuetal2018} , as well as newly identified streams by \citet{ibataetal2018}.

Out of a family of $\sim$160 Galactic globular clusters, nearly 50 have been associated with
one (or several) of the accreted galaxies and tidal streams discovered so far in the halo, based on their
projected positions, kinematics and chemical abundances \citep{bellazzinietal2003,fb2010,massarietal2019}. However, the unambiguous confirmation of the
extra-Galactic origin of one of the halo globular clusters requires the combination of multiple techniques
and an important observational effort \citep[see][]{carballobelloetal2017}, so that most of the globular
clusters on that list remain as candidates.

\begin{figure}
\includegraphics[width=\columnwidth]{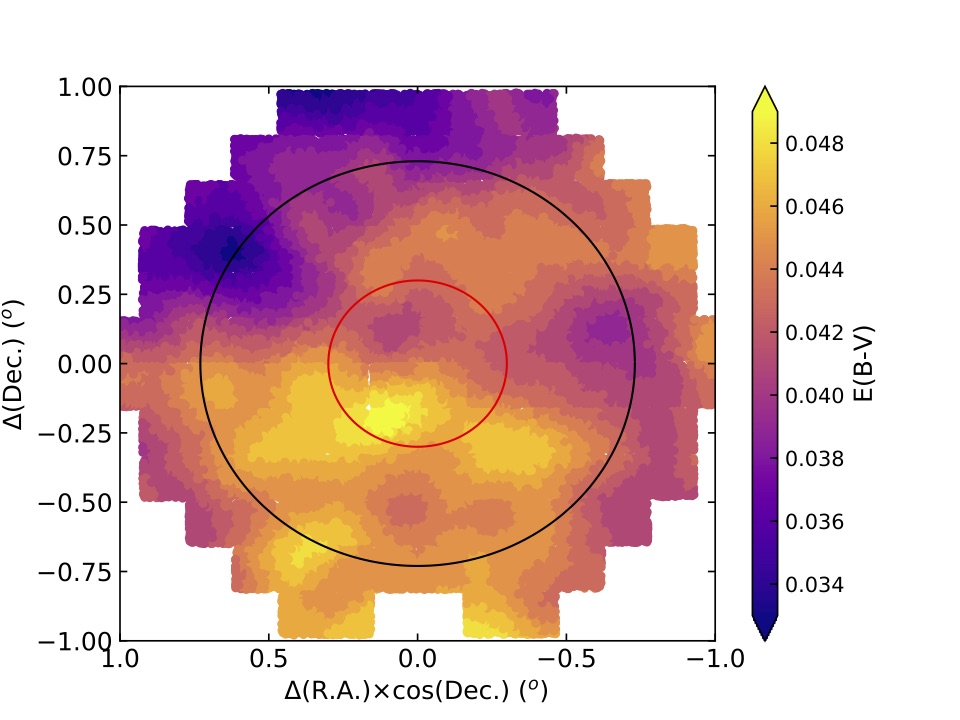}
\caption{Reddening map across the field of NGC\,7099. The red and black circles are of 0.32$\degr$ 
and 0.73$\degr$ in radius, respectively.  The black circle separates two areas of the same size.}
\label{fig:fig1}
\end{figure}

\begin{figure}
\includegraphics[width=\columnwidth]{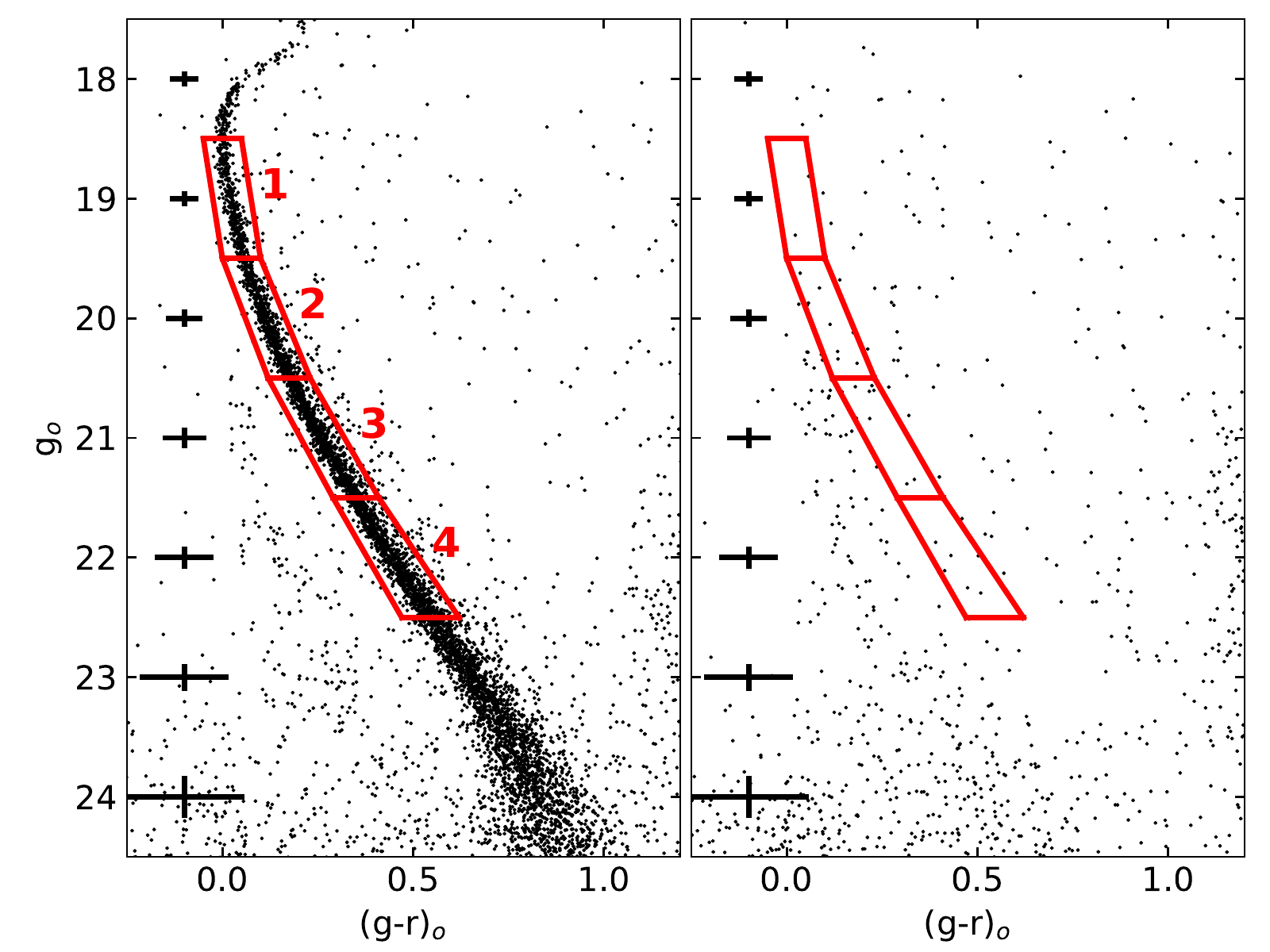}
\caption{CMDs of stars in the field of NGC\,7099 ($r$ $<$ 0.15$\degr$; left panel) and in an annular region of same area centered on the cluster with an external radius of 0.8$\degr$ 
(right panel). The four regions along the cluster MS used to perform star counts are delineated with red contour lines.}
\label{fig:fig2}
\end{figure}

It has been thought that the structure of Milky Way globular clusters might provide precious information
about their origin and accretion  processes, as well as about the cluster internal 
dynamical evolution. Interestingly, many of the hypothetically accreted
halo globular clusters show the signature of the presence of extra-tidal stars in the form of tidal tails 
\citep[e.g., Pal\,5,][]{odenetal2003} and extended halos \citep[e.g., 47 Tuc,][]{p17c}. For instance, in the case of the Sausage-Enceladus accretion event, most of the associated globular clusters proposed by \citet{myeongetal2018} are embedded in a stellar component,
which is only revealed using matched-filter techniques on deep wide-field photometry 
\citep{carballobelloetal2014,vanderbekeetal2015,kuzmaetal2016,carballobelloetal2018,carballobello2019,pc2019}.
Whether the outer structure of a globular cluster is defined by its formation conditions within a smaller
protogalactic fragment or exclusively by its internal evolution and the orbit followed around the Milky Way, is one of the open questions in the study of Milky Way globular clusters  \citep{massarietal2019,pcb2020}.

In this study we focus on NGC\,7099, a poorly studied Milky Way globular cluster that might have
been formed within the Sausage-Enceladus galaxy \citep{massarietal2019}. With peri and apogalactocentric
distances of 1.49 kpc and 8.15 kpc, respectively \citep{baumgardtetal2019}, NGC\,7099 moves along its retrograde
orbit with a relatively high eccentricity and inclination angle \citep[$e$= 0.69, $i$=61.5$\degr$,][]{piatti2019},
so that it could have crossed many times the inner disk regions. NGC\,7099 has lost $\sim$ 30$\%$ of 
its initial mass by tidal heating caused by the Milky Way’s gravitational field, in addition to another half 
of its initial mass by stellar evolution \citep{piattietal2019b}. In Section 2, we describe the 
observational data collected, on which our analysis relies. Sections 2 and 3 detail the cleaning 
procedure applied to the observed NGC\,7099 color-magnitude diagram (CMD) to get rid of
the field star contamination, build the respective stellar density maps, and discuss our results in the 
light of recent findings \citep[][hereafter S20]{sollima2020}.

\section{Data collection and processing}

In this work we have used the Dark Energy Camera (DECam), which is mounted at the prime focus of the 4-m Blanco 
telescope at Cerro Tololo Inter-American Observatory (CTIO). DECam provides a 3\,deg$^{2}$ field 
of view with its 62 identical chips with a scale of 0.263\,arcsec\,pixel$^{-1}$ \citep{flaugheretal2015}. 
NGC\,7099 was included among the targets of the observing program 2019B-1003 (08/09.08.2019) 
and the exposure times were 600$\times$4\,s and 600$\times$4\,s, for the $g$ and $r$ bands, 
respectively. We also observed 3--5 SDSS fields per night at different airmass to derive the 
atmospheric extinction coefficients  and the transformations between the instrumental magnitudes 
and the SDSS $ugriz$ system \citep{fukugitaetal1996}.

The images were processed by the DECam Community Pipeline \citep{valdesetal2014} and accessed 
via the NOAO Science Archive. The photometry was obtained from the  images with the PSF-fitting
 algorithm of \textsc{daophot\,ii/allstar}
\citep{s1987}. The final catalog only includes stellar-shaped objects with $|sharpness| \leq  
0.5$ to avoid as much as possible the presence of background galaxies and non-stellar sources in 
our analysis. 

We also used {\sc daophot\,ii} to include in our images synthetic stars in order to estimate the completeness of our photometric catalogs. After applying our photometry pipeline, the limiting magnitude was set at the magnitude at which the synthetic stars are only recovered in 50\% of the altered images. For $g$ and $r$ bands, the limiting magnitudes in our catalogs are 23.4\,mag and 23.3\,mag, respectively.

\begin{figure*}
\includegraphics[width=\columnwidth]{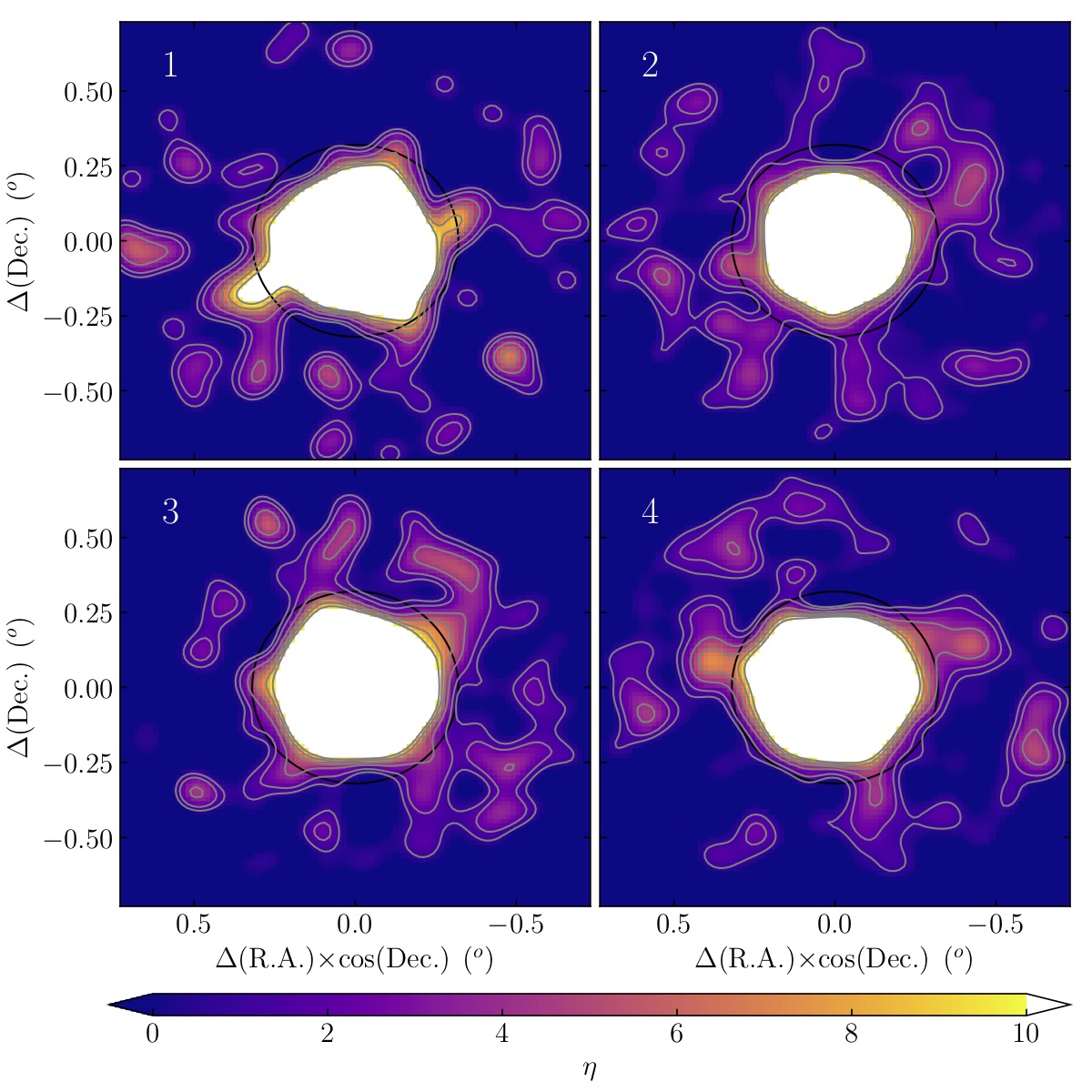}
\includegraphics[width=\columnwidth]{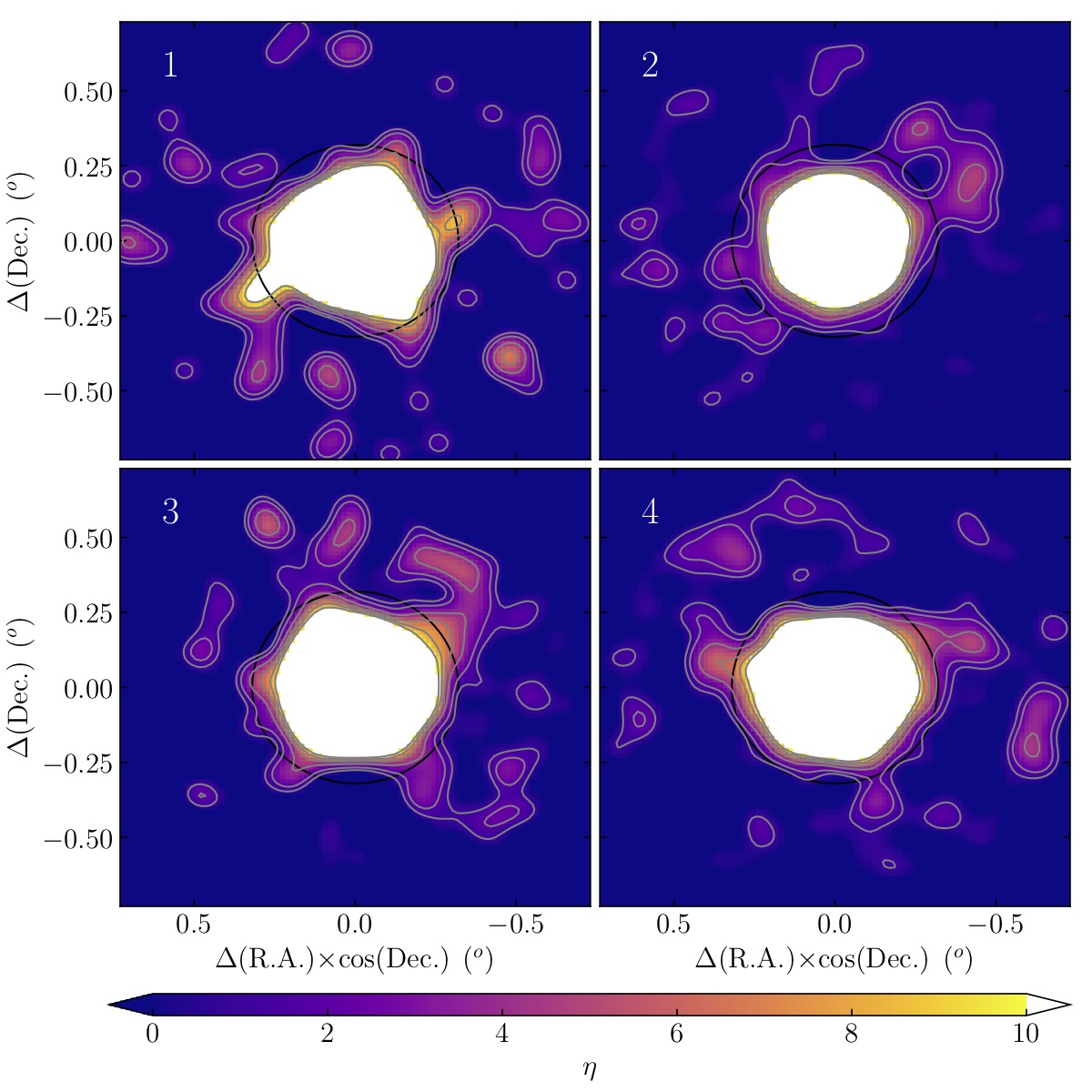}
\caption{Observed (left panels) and field star cleaned (right panels) stellar density maps for the four cluster MS strips of Figure~\ref{fig:fig2} as
labelled at the top-left margin of each panel. The black circle centered on the cluster indicates the 
assumed tidal radius. Contours for $\eta$ = 1, 2, 4, 6, 8 and 10 are also shown; the colors follow 
the coding shown at the bottom of each panel.}
\label{fig:fig3}
\end{figure*}

\begin{figure}
\includegraphics[width=\columnwidth]{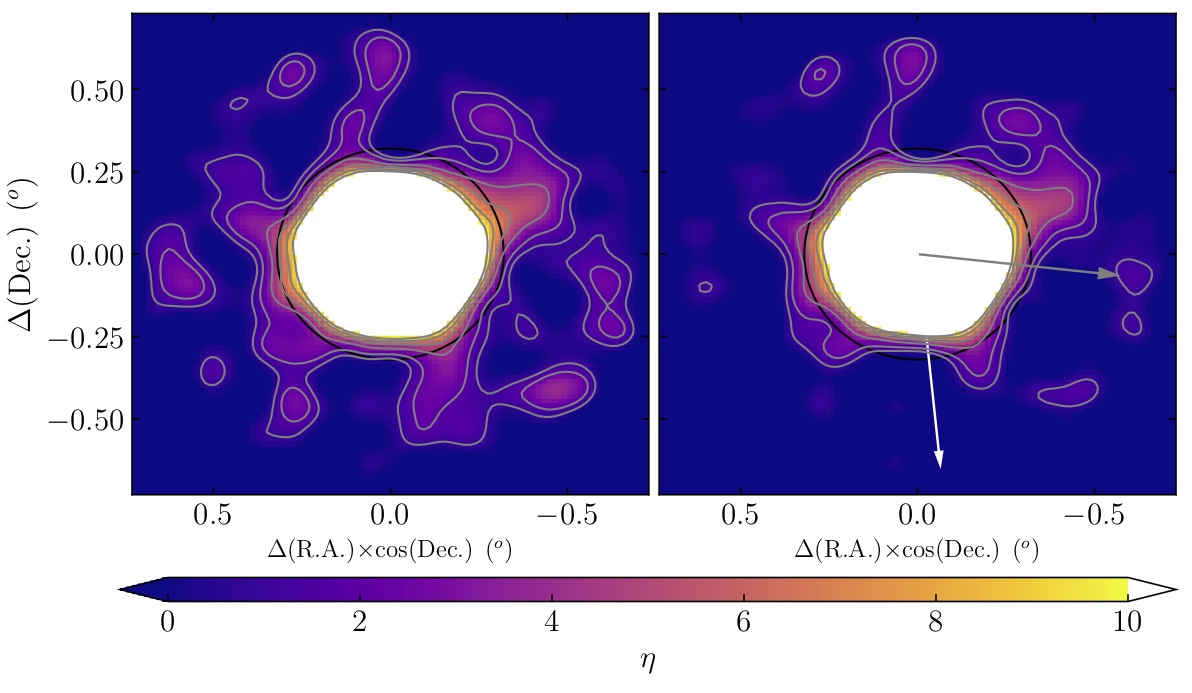}
\caption{Observed (left panel) and field star cleaned (right panel) stellar density maps} adding up all stars in the four cluster MS strips of 
Figure~\ref{fig:fig2}. The black circle centered on the cluster 
indicates the assumed tidal radius. Contours for $\eta$ = 1, 2, 4, 6, 8 and 10 are also shown.
The different arrows indicate the directions of the cluster proper motion (white) and of the 
Milky Way center (grey). 
\label{fig:fig4}
\end{figure}

\begin{figure}
\includegraphics[width=\columnwidth]{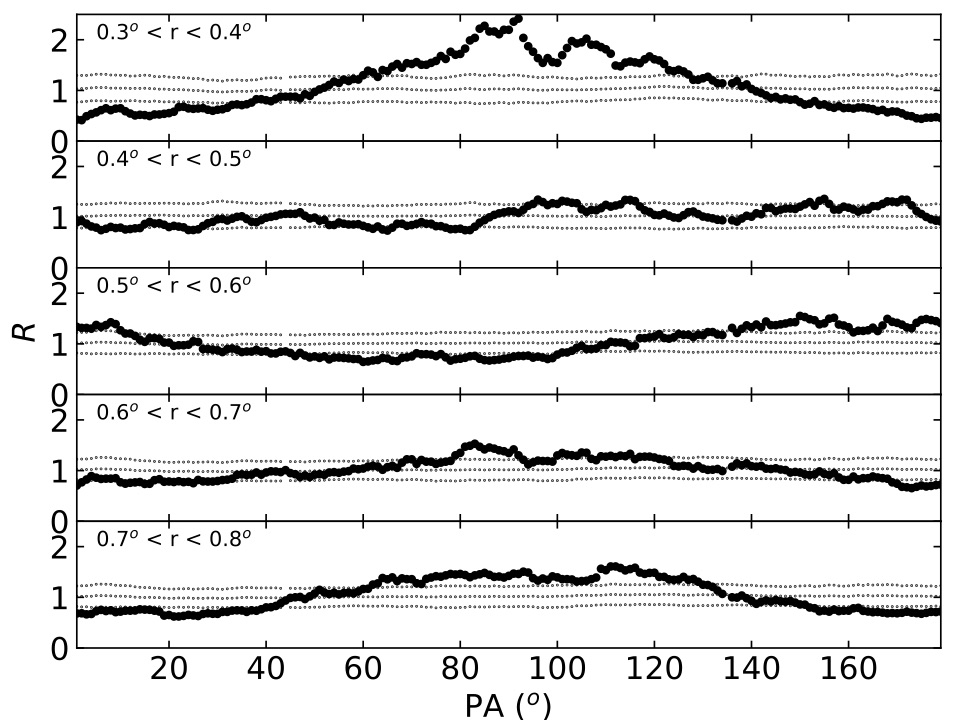}
\caption{The ratio $R$= $(N^A_c N^B_f)/(N^B_c N^A_f)$ (see Section 3 for details)
 versus PA  obtained from star counts in the observed MS strip (large filled circles). Dotted lines represent the resulting mean and dispersion of the Monte Carlo simulations. The inset label
 indicates the annular region considered.}
\label{fig:fig5}
\end{figure}

We retrieved the $E(B-V)$ values as a function of R.A. and Dec. from \citet{sf11} provided by 
NASA/IPAC Infrared Science Archive\footnote{https://irsa.ipac.caltech.edu/} for the entire analyzed 
area. Figure~\ref{fig:fig1} shows that the interstellar extinction along the line-of-sight is low, with a
maximum difference between the most and least reddened regions of $\la$ 0.02 mag. Using
Figure~\ref{fig:fig1} we assigned individual $E(B-V)$ values to the measured stars according to
their positions in the sky. In order to correct the observed magnitudes and colors by interstellar 
extinction, we used the individual $E(B-V)$ values and the $A_\lambda/A_V$ coefficients given by 
\citet{wch2019}. Aiming at illustrating the wealth of information we gathered, Figure~\ref{fig:fig2}
depicts the intrinsic CMD of the inner cluster region ($r$ < 0.15$\degr$) and that of a sky annular
region centered on the cluster with equivalent area and external radius of 0.8$\degr$. By comparing
both panels of Figure~\ref{fig:fig2} we conclude that the NGC\,7099 CMD is affected by a low
field contamination. Figure~\ref{fig:fig2} also reveals a relatively narrow cluster MS, which extends 
more than 5 mag in $g$.

\section{Stellar density maps}

Because of two-body relaxation, less massive stars are candidates to escape the cluster more
easily than more-massive ones \citep{bg2018}. Hence, fainter cluster Main Sequence (MS) stars are usually employed
to search for extra-tidal structures around globular clusters 
\citep[see][and references therein]{pft2020}.  We therefore take advantage the
well-delineated MS shown in Figure~\ref{fig:fig2} to define four adjacent 
strips, from underneath the MS turnoff down to 4 mag below it, to map the distribution of their stellar 
populations. The cluster stellar density
map is then straightforwardly constructed for the four different subsets of cluster MS stars.

Figure~\ref{fig:fig3} shows the stellar density maps built for each MS strip using the AstroML 
\citep{astroml} kernel density estimator (KDE) routine. We superimposed a grid of 100$\times$100 
square cells on to the area of interest ($r$ $<$ 0.73$\degr$) and used a range of values for the KDE
bandwidth from 0.02$\degr$ up to 0.07$\degr$ in steps of 0.01$\degr$ in order to apply the KDE to 
each generated cell. We adopted a bandwidth of 0.05$\degr$ as the optimal value.
We also estimated the background level using the stars distributed within the annular region defined
by the black circle in Figure~\ref{fig:fig1} and an external radius of 0.8$\degr$. We divided such an
annulus into 16 adjacent sectors of 22.5$\degr$ wide, and counted the number of stars inside them.
We rotated such an array of sectors by 11.25$\degr$ and repeated the star counting. Finally, we 
derived the  mean  value in the 32 defined sectors, which turned out to be 35.5, 133.3, 114.0,
and 100.0 stars/deg$^2$ for the MS strips \#1 to 4, respectively. As for the standard deviation,  we
performed a thousand Monte Carlo realizations using the stars located beyond the red circle
in Figure~\ref{fig:fig1}, which were rotated randomly (one different angle for each star) before 
recomputing the density map. The resulting standard deviation of all the generated density maps 
turned out to be 20.3, 47.7, 41.4, and 44.6 stars/deg$^2$ for the MS strips \#1 to 4, respectively.
The color scale in Figure~\ref{fig:fig3} represents the absolute deviation from the mean value in the
field, in units of the standard deviation, that is,  $\eta$ = (signal $-$ mean value)/standard deviation. 
We have painted white stellar densities with $\eta$ $>$ 10 in order to highlight the least dense structures. The same Monte Carlo procedure described above has been employed to determine the background density and its standard deviation in the cleaned CMD and have been used to calculate the value of $\eta$.
 
We applied a procedure to get rid of field stars that fall inside the four defined MS strips. The method
was devised by \citet{pb12} and used satisfactorily for cleaning CMDs of star clusters projected towards
crowded star fields  \citep[e.g.,][and references therein]{p17a,p17b,p17c} and affected by differential 
reddening \citep[e.g.,][and references therein]{p2018,petal2018}. The method uses the magnitudes
and colors of stars in a reference field CMD to subtract an equal number of stars in the cluster
CMD that best resembles the reference field CMD. This is done by considering the position of each
field star in the cluster CMD and by subtracting the closest star in the cluster CMD to that field star. 
The star-to-star subtraction in the cluster CMD is necessary to avoid stochastic
effects and subtraction residuals, which arise when any fixed size for boxes homogeneously
distributed throughout the CMD is chosen to count stars into them. We thus tightly reproduce
the reference field CMD in terms of luminosity function, color distribution, and stellar density.
The purpose of such a cleaning is to obtain a cluster CMD without field contamination. 
The contribution of field stars is included in the observed stellar density maps.
The cleaning procedure subtracts exactly the number of field stars that contribute to the
reference field CMD. The procedure searches throughout the whole cluster
area for field stars to eliminate, giving the same chance (weight) to every subregion
to contain such a field star. This is done to avoid spurious overdensities in the cleaned cluster field. 
 If the star field were homogeneous and contained $N$ stars per 
unit area, the procedure would subtract $N$ stars per unit area.
If there were any intrinsic spatial gradient of field stars in the cluster area, the procedure would eliminate it,  at the expense of leaving some residuals as a counterpart of the excess
of stars subtracted from the inner cluster region. If we did not considered the inner cluster region,
the residuals would certainly decrease.
In practice, for each reference field star, we randomly select an annular subregion in the
cluster area where to subtract a field star. If no star in found in that annular sector,
we randomly select another one and repeat the search, and allow to iterate the procedure 
up to 1000 times. The annular regions are 90$\degr$ wide
and of constant area. Their external radii are chosen
randomly, while the internal ones are calculated so that the areas of the annular
sectors are constant. Here we adopted an area equal to $\pi$$r_{cls}$, where $r_{cls}$
is the cluster radius.
If any stellar feature remains in the density map built 
from the cleaned cluster CMD, that is assumed to be an intrinsic cluster feature. 
The stellar density map built from stars in that  CMD should have a background density close to zero.
In doing this, we considered the uncertainties in magnitudes and colors by repeating the procedure 
hundreds of times with magnitudes and colors varying within their respective errors.  For the designed
MS strips, photometric errors increase from $\approx$ 0.04 mag up to 0.09 in $g_0$ and from 
$\approx$ 0.03 mag up to 0.11 mag in $(g-r)_0$ for the range $g_0$ $\approx$  18- 23 mag.  As for the 
reference star field, we chose the region embraced by the DECam boundaries and the black circle
in Figure~\ref{fig:fig1}. We chose that region with the aim of dealing as close as possible with the 
intrinsic star field characteristics in the direction toward the cluster.  Its stellar density does
not show any dependence as a function of the position angle for any of the four MS strips.

The field star cleaned stellar density maps were built similarly as the observed ones, and are
shown in Figure~\ref{fig:fig3}. They reveal a short tail, which resembles that of S20 for the same
area, and some stellar debris distributed beyond the tidal radius
of NGC\,7099  \citep[0.32$\degr$][]{harris1996,piattietal2019b}.  Although the MS strips \#1 to 4
include stars with different masses, in the sense that the fainter the MS strip the less massive its
stars, we do not find any remarkable difference between density maps. We would have expected 
more extended extra-tidal features to show up in the case of the lowest-mass bins, as lower-mass
stars can be more easily stripped away from the cluster than their higher-mass counterparts.
For this reason, we also produced a stellar
density map with all the stars in the MS strips \#1 to 4. Figure~\ref{fig:fig4} depicts the
resulting stellar density maps, where evidence of a short tidal tail would seem to arise clearer.

\begin{figure}
\includegraphics[width=\columnwidth]{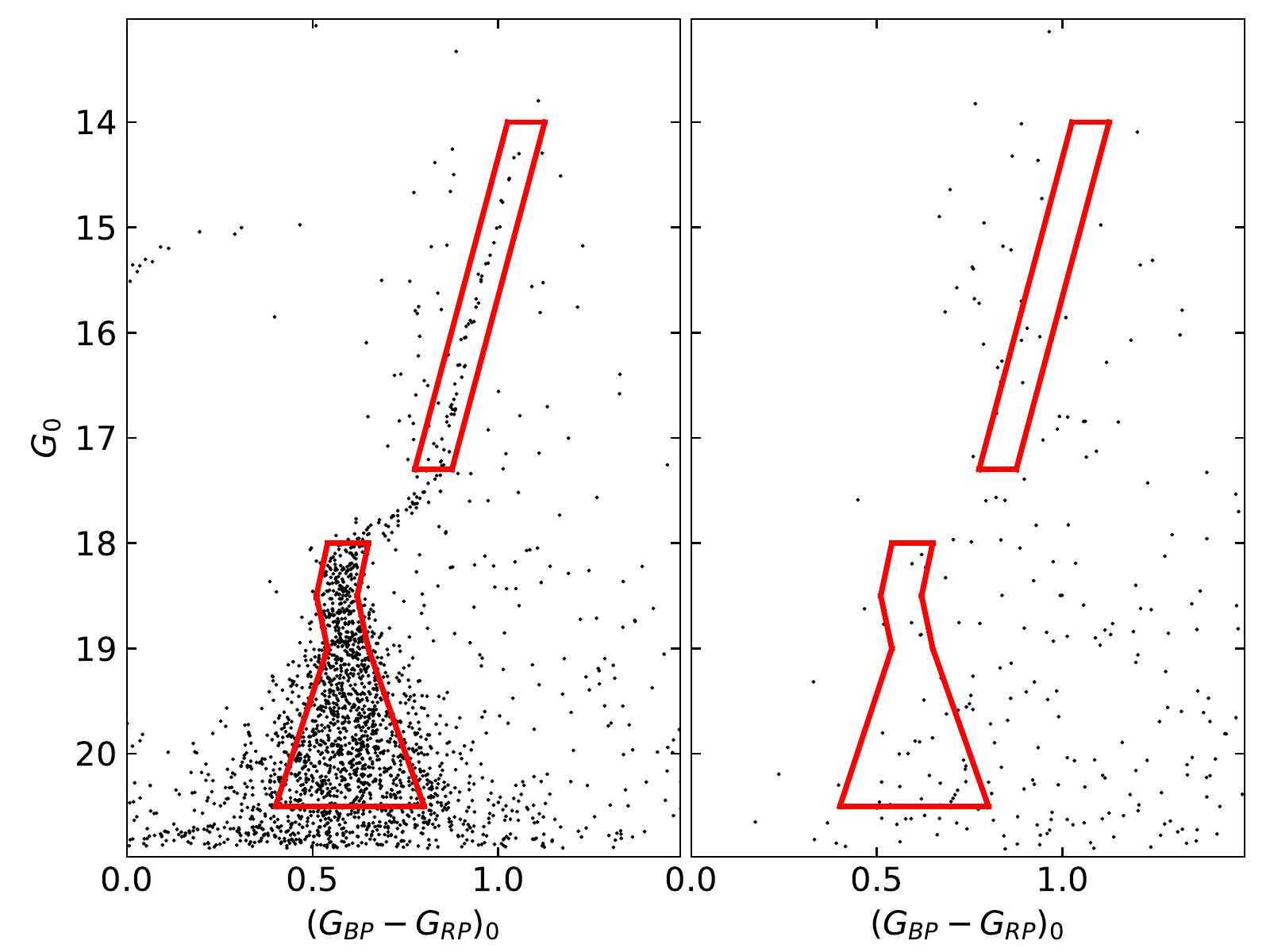}
\caption{{\it Gaia} color–magnitude diagrams of stars in the field of NGC\,7099 ($r$ $<$ 0.15$\degr$; left panel) and in an annular region of same area centered on the cluster with an external radius of 0.8$\degr$ 
(right panel). The RGB and MS regions used to perform star counts are delineated with red contour lines.}
\label{fig:fig6}
\end{figure}

\begin{figure}
\includegraphics[width=\columnwidth]{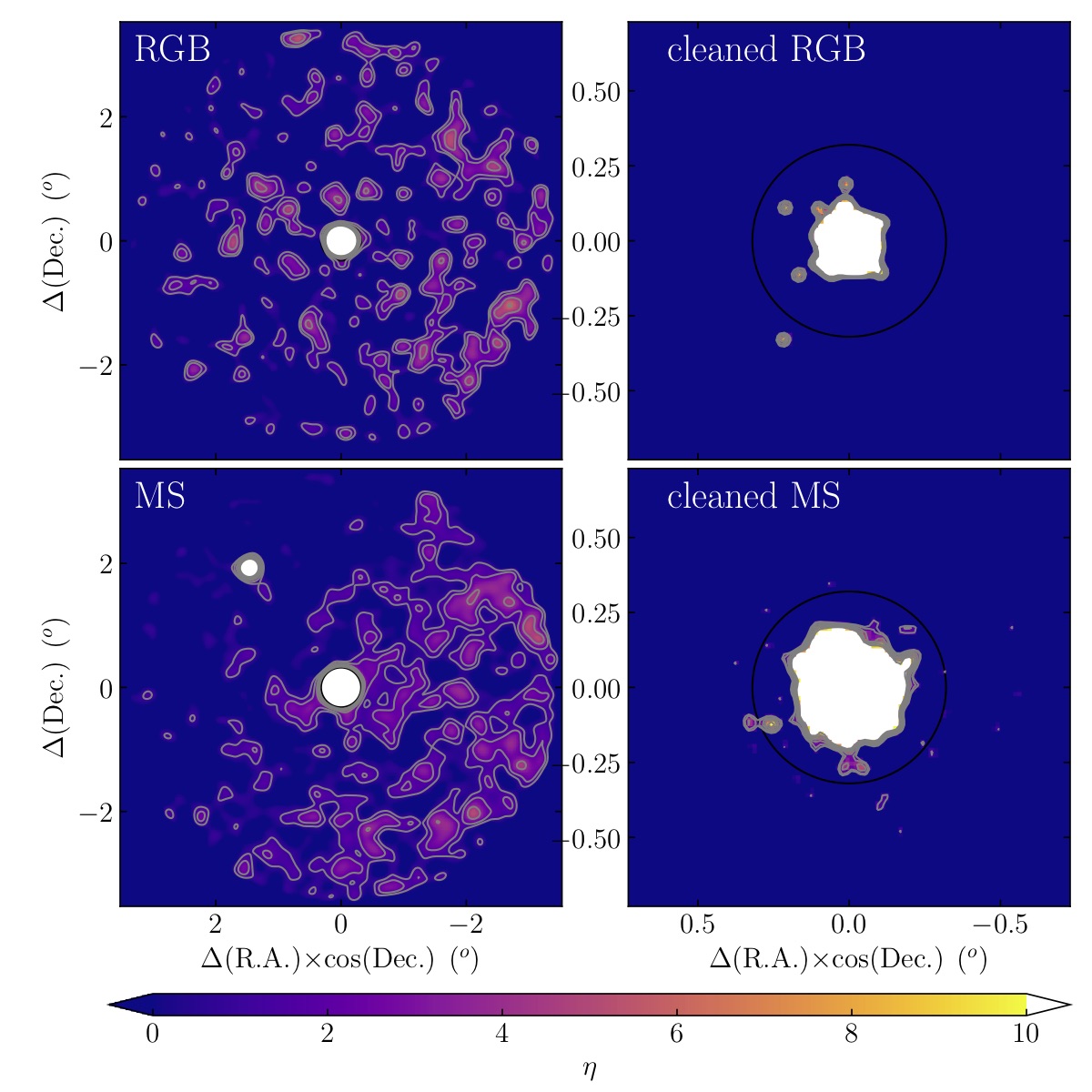}
\caption{{\it Gaia} observed (left panels) and field star cleaned (right panels) stellar density maps for the 
cluster RGB and MS strips of Figure~\ref{fig:fig6}. The black circle centered on the cluster indicates the 
assumed tidal radius.}
\label{fig:fig7}
\end{figure}

We further analyzed the possibility of tidal deformations across the cluster stellar
density map, in the sense that preferential orientation toward the Galactic center
and along the direction of the  orbit of the cluster are expected in the innermost and
outermost parts, respectively \citep{montuorietal2007}. We followed the recipe applied by
\citet{sollimaetal2011} based on  counts of cluster MS strip stars in alternate
pairs of circular sectors of  90$\degr$ in width located at a given distance from the
cluster center and oriented at a position angle (PA) in opposite directions. 
PA is measured from the North anti-clockwise. We then computed the ratio $R(PA) = (N^A_c N^B_f)/(N^B_c N^A_f)$, where $A$ and $B$
are the pair of alternate sectors, and $c$ and $f$ refer to the cluster MS strip
and a CMD field rectangle defined by 18.5 $<$ $g_0$ (mag) $<$ 22.5 and 
1.0 < $(g-r)_0$ (mag) < 1.15. In order to assess the statistical significance of our
results, we performed 1000 Monte Carlo realizations using the same number of
measured stars distributed randomly in PA and then obtained the mean and
standard deviations of those independent executions. Figure~\ref{fig:fig5} depicts the
 resulting curves for different annular regions. As suggested by Figure~\ref{fig:fig4},
 Figure~\ref{fig:fig5} shows the existence of a short tidal tail nearly aligned with direction to the
 Milky Way center, in very good agreement with \citet{montuorietal2007}'s models.
 The PA's width of this short tail also agrees with S20 (see his Figure 4).
For completeness purposes we included in Fig.~\ref{fig:fig4}
 (right panel) the directions toward the Milky Way center and of the 
 motion of the cluster. 
 
We estimated the $g$ surface brightness of the tidal tail observed in Fig.~\ref{fig:fig4},
i.e., from the stars that were not subtracted by the field star cleaned CMD procedure, for three different adjacent annulus
of 0.1$\degr$ wide, located between 0.4$\degr$ and 0.7$\degr$ from the cluster center.
We first calculated the integrated $g$ magnitude ($g_{int}$) for stars located inside a radius 
of 0.1$\degr$ from the cluster center in Fig.~\ref{fig:fig4} (right panel), by summing de DECam $g$
fluxes of individual stars. A normalization constant $c$ was calculated as :
  
$ c = {g_{int}}_0 + 2.5\times log(N_0/\pi 0.1^2)$
  
\noindent where $N_0$ represents the number of stars used. Then, the integrated magnitudes
for the different annular regions were calculated according to:

$g_{int} = c - 2.5\times log(N/(\pi(r_2^2 - r_1^2))) $

\noindent where $N$, $r_1$, and $r_2$ represent the number of stars located in an annulus
with inner and outer radii $r_1$ and $r_2$, respectively. With the aim of expressing the
integrated magnitudes in units of $V$ mag per square arcsec, we used the integrated $B-V$
color for NGC\,7099 listed by  \citet[=0.60 mag;][2010 Edition]{harris1996} and the linear
relationship between $g$ and $V$ in terms of the $B-V$ color \citep{jetal06}.
 The resulting surface brightness and its uncertainty computed from propagation of errors
  assuming a Poisson statistics turned out to be 31.35$\pm$0.11, 31.41$\pm$0.11, and 31.44$
 \pm$0.12 mag per square arcsec for the annulus at $r$ = 0.4$\degr$-0.5$\degr$, 
 0.5$\degr$-0.6$\degr$, and 0.6$\degr$-0.7$\degr$, respectively.

\section{Discussion}

Recently,  S20 presented results of a 5D mixture modelling technique
based on {\it Gaia} DR2 data of the outer regions of 18 Milky Way globular clusters. He found that 
NGC\,7099 has long tidal tails. Reportedly, those tails are composed by cluster red giant branch (RGB)
and MS stars reaching down to $\sim$ 1 mag below the MS turnoff of the cluster. This means that S20
used stars on average brighter than those of our selected MS strip \#1 in Figure~\ref{fig:fig2}. 
At this point, we wonder what caused  the different results obtained in the present study compared to 
those in S20, particularly because MS stars in strips \#2 to 4 are expected to be better candidates to 
trace extra-tidal structures. According to S20 (his Figure 1), selected MS stars have parallaxes
within a range at least 10 times larger than the parallaxes of RGB stars, which means that 
field stars with loci in the CMD and in the vector point diagram (VPD, proper motion in R.A. (pmra) 
vs. proper motion in Dec. (pmdec)) similar to those of the
cluster stars could be included in  the selected sample. From a comparison of previous studies on the
outer regions of  Milky Way globular clusters \citep[see Table 1 in][]{pcb2020}, we also found that 
results for some globular clusters are distinct from those found in S20. For instance, NGC\,1851 and
4590 are known to have long tidal tails, although not detected in S20, while NGC\, 2298, 6341, and
6362  seem not to have the tails traced in S20.  To this respect, we note that it is not straightforward to conclude
whether deep photometric-only data sets or shallow proper motion-selected ones are
preferable to detect tidal tails. Therefore,  it is conceivable that differences can be found
in some globular clusters subject to different levels of contaminations when analyzing them with 
different data sets. 

We used the 5$\degr$ in radius around NGC\,7099  {\it Gaia} DR2 \citep{gaiaetal2016,gaiaetal2018b}
database as in S20 to repeat the CMD cleaning procedure applied in Section\,3, in order to
build a stellar density map to be compared with that produced by S20. Figure~\ref{fig:fig6} shows
the intrinsic CMD of the inner cluster region used in Figure~\ref{fig:fig2} that highlights the main
features of NGC\,7099. Particularly, we chose a strip along the RGB and another one
from the MS turnoff down to  2.5 mags underneath. According to \citet[][see Section 3]{arenouetal2018},
the {\it Gaia} DR2 photometry completeness is 90$\%$ for stars with $G <$ 19 mag in the inner region
of a globular cluster with $\sim$10$^4$ stars/sq deg, so that we deal with basically a complete 
photometry data set. The adopted fainter limit should be likewise aceptable to guarantee the
homogeneity of the completeness. We also used a bandwidth of 0.1$\degr$ in order to get
the highest contrast of tidal tails.
We cleaned both strips from field star contamination for a circular area
centered on NGC\,7099 with radius equal to 3.53$\degr$, in order to ensure a reference field
star region (3.53$\degr$ $<$ $r$ $<$ 5$\degr$) with an equal cluster area. The {\it Gaia}
photometry was corrected by reddening  using individual $E(B-V)$ values obtained from \citet{sf11},
and the relationships $A_G$ = 2.44$E(B-V)$ and $E(G_{BP}-G_{RP})$ = 1.27$E(B-V)$ 
\citep{wch2019}. The observed and cleaned stellar density maps are depicted
in Figure~\ref{fig:fig7}.  Observed RGB and  MS density maps exhibit different spatial
distributions, the latter suggesting the existence of tidal tails within a remarkably non-uniform
star field spatial distribution. Long tidal tails do not remain in the star field cleaned density maps,
although some hint for short ones are found.

\begin{figure}
\includegraphics[width=\columnwidth]{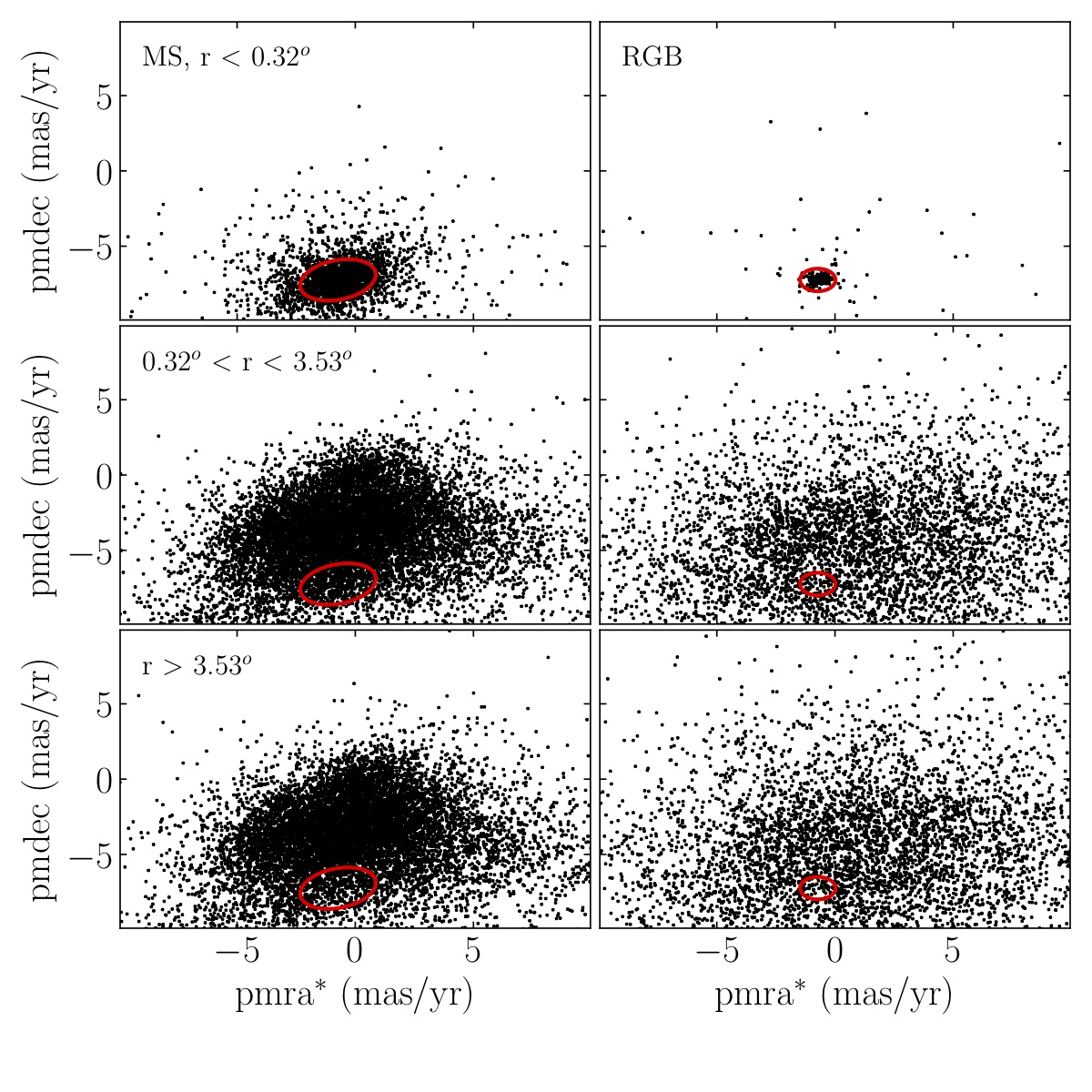}
\caption{Vector point diagrams for RGB (right panels) and MS (left panels) strips of 
Figure~\ref{fig:fig6} for different annular regions. Red ellipses that mostly embrace cluster stars
in the top panels were superimposed onto the middle and bottom panels.}
\label{fig:fig8}
\end{figure}

The above outcomes suggest that stars distributed along the cluster CMD features (e.g., strips in
Figures~\ref{fig:fig2} and \ref{fig:fig6}) located beyond its tidal radius ($r >$ 0.32$\degr$)  might
belong to  the Milky Way field.  We further used the proper motions of  all the stars located
within the strips drawn in Fig.~\ref{fig:fig6}, which show 
some particular features. We illustrate them in Figure~\ref{fig:fig8}, where we plotted the VPD for 
RGB and MS strip stars located in three different sky regions, namely, those located inside the tidal 
radius of the cluster (top panels), those distributed between 0.32$\degr$ and 3.53$\degr$ from the 
cluster center (middle panels), and those adopted as reference field stars (bottom panels). Cluster 
stars  ($g_0$(MS strip) $<$ 19 mag) and RGB strip stars) have a well-defined distribution in the 
VPD, which we  embraced 
by a red ellipse (top panels). Stars located beyond the cluster radius have a different distribution 
(middle and bottom panels), with some superposition with that of cluster stars (top panels). 
For comparison purposes, we superimposed the ellipses in the top panels onto the middle 
and bottom panels. The middle panel
reveals that there are many MS strip stars located outside the cluster with proper motions distinguishable from that of the cluster  (most of the stars located outside the ellipses). 
Particularly, they have $g_0$ mag between 19 and
20.5 mag  (see Figure~\ref{fig:fig6}). Most of those stars were eliminated using the CMD
cleaning procedure (see Fig.~\ref{fig:fig7}).

\begin{figure}
\includegraphics[width=\columnwidth]{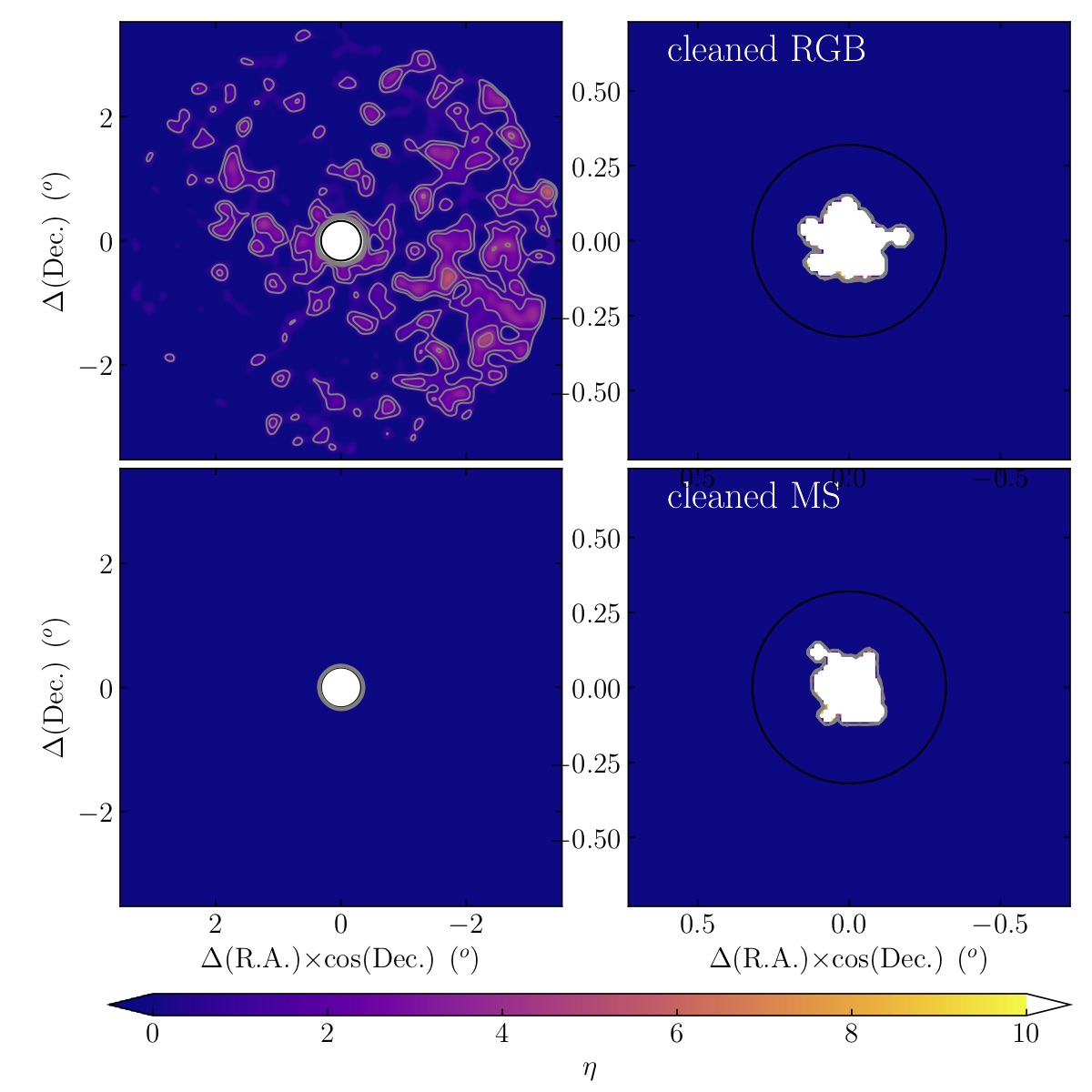}
\caption{{\it Gaia} observed (top-left panel) and field star cleaned (bottom-left panel) stellar density 
maps for stars selected only from proper motion and parallax criteria. The right panels correspond
to RGB and MS stars that remained unsubtracted after the photometric and kinematic
cleaning procedures were applied.The black circle centered on the cluster indicates the 
assumed tidal radius.}
\label{fig:fig9}
\end{figure}

From Figure~\ref{fig:fig8} we conclude that out of all the stars distributed within the RGB and MS strips,
some of them have sky projected kinematics similar to that of cluster stars. Therefore, an
unavoidable question arises: are those stars cluster members? If we used the CMD cleaning
procedure, we would answer that they do not belong to the cluster population, as obtained above.
Indeed, we clean from field star contamination the VPD at $r$ $<$ 3.53$\degr$ from
the cluster center, using as a reference  a star field  located between 3.53$\degr$ and 5$\degr$, similarly as we carried out for the CMD cleaning. The resulting star field cleaned stellar density 
maps for RGB and MS strip stars that share the cluster's proper motion, built using the same 
bandwidth as in Figure~\ref{fig:fig7}, do not show any extra-tidal feature  (see Fig.~\ref{fig:fig9},
right panels).  Finally, we used only the kinematic  and parallax information for selecting cluster stars. The top-left panel 
of Fig.~\ref{fig:fig9} shows the stellar density map built for stars with proper motions
within an ellipse centered on (-0.73 mas/yr,-7.24 mas/yr), width=3.4 mas/yr, height=2.5 mas/yr and
angle=30$\degr$, and  $|\varpi|$ $<$ 5$\sigma(\varpi)$ (S20), while the bottom-left one
depicts the stellar density map after cleaning the cluster field from field contamination, 
based only on the kinematic information. As can be seen, there is not any remaining signature
of extra-tidal features.

\section{Conclusions}

In this study we analyze the outer regions of the Milky Way globular cluster NGC\,7099, aiming at
identifying extra-tidal features. The cluster caught our attention because it was claimed that
it might have formed within the accreted Sausage-Enceladus dwarf galaxy and deposited in an inner halo orbit,
with relatively high eccentricity and inclination angle with respect to the Milky Way plane. According to
\citet{piattietal2019b}, globular clusters with orbital parameters like those of NGC\,7099 have lost
relatively more mass by tidal disruption than globular clusters rotating in the Milky Way disk. In the 
case of NGC\,7099, the disrupted mass to the initial mass ratio is 0.30.

We carried out DECam observations that, as far as we are aware, allowed us to build the deepest
cluster CMD, reaching $\sim$ 6 mags beneath its MS turnoff. Those observed faint MS
stars are treated as suitable candidates to trace extra-tidal features, because they are the first to
cross the Jacobi radius once they reach the cluster boundary driven by two-body relaxation.
Indeed, it has long been observed that the brighter the range of magnitudes considered, the sharper 
the cluster stellar radial profile. 

In order to monitor any differential change in the stellar density map caused by cluster stars
with distinct brightnesses, we split the long cluster MS into four segments of one mag long each, and
analyzed them separately. We built their respective stellar density maps with a frequently used KDE
technique, once the magnitudes and colors of the stars were individually corrected by the interstellar 
extinction. When a statistical decontamination method is used to clean the cluster CMD from field 
stars, the resulting cleaned stellar density maps show  a short tidal tail and
some scattered  debris. The short tidal tail is nearly oriented to the Milky Way center, 
as expected for innermost parts of tidal tails. We did not found these extra-tidal features when 
analyzing {\it Gaia} data with the same cleaning procedure. In this case, the photometry used is 
much shallower than the DECam photometry, reaching 2.5 magnitudes below the 
cluster MS turnoff, and the analyzed area 25 times larger than afforded by our DECam data,  so that the reference star field is located much farther from the cluster 
(3.53$\degr$ $<$ $r$ $<$ 5.00$\degr$).

RGB and MS strip stars located inside and outside the cluster tidal radius (0.32$\degr$) present distinguishable VPDs, with some superposition. When comparing  their VPDs, 
using the CMD cleaning method separately, for stars located within a circle of radius 3.53$\degr$ 
with those placed in the reference star field (3.53$\degr$ $<$ $r$ $<$ 5.00$\degr$), we did not find 
any extra-tidal feature. S20 detected long tidal tails, the innermost parts of them also
found from DECam data. Finally, we would like to mention that extra-tidal stars should 
have distinct velocities that make them possible to escape the cluster.  Likewise, they are at different 
Galactocentric distances than the cluster itself, so that they are differently affected by the Milky Way gravitational field \citep{piatti2020,dk2020}.

\begin{acknowledgements}
We thank the referee for the thorough reading of the manuscript and
timely suggestions to improve it.

Support for M.C. is provided by ANID Millennium Science Initiative grant ICN12\textunderscore009; by Proyecto Basal AFB-170002; and by FONDECYT grant 1171273.

Based on observations at Cerro Tololo Inter-American Observatory, NSF’s NOIRLab (Prop. ID 2019B-1003; 
PI: Carballo-Bello), which is managed by the Association of Universities for Research in Astronomy (AURA)
under a cooperative agreement with the National Science Foundation.

This project used data obtained with the Dark Energy Camera (DECam), which was constructed by the 
Dark Energy Survey (DES) collaboration. Funding for the DES Projects has been provided by the US 
Department of Energy, the US National Science Foundation, the Ministry of Science and Education of Spain, 
the Science and Technology Facilities Council of the United Kingdom, the Higher Education Funding Council 
for England, the National Center for Supercomputing Applications at the University of Illinois at 
Urbana-Champaign, the Kavli Institute for Cosmological Physics at the University of Chicago, Center for 
Cosmology and Astro-Particle Physics at the Ohio State University, the Mitchell Institute for Fundamental 
Physics and Astronomy at Texas A\&M University, Financiadora de Estudos e Projetos, Funda\c{c}\~{a}o 
Carlos Chagas Filho de Amparo \`{a} Pesquisa do Estado do Rio de Janeiro, Conselho Nacional de 
Desenvolvimento Cient\'{\i}fico e Tecnol\'ogico and the Minist\'erio da Ci\^{e}ncia, Tecnologia e Inova\c{c}\~{a}o, the Deutsche Forschungsgemeinschaft and the Collaborating Institutions in the Dark Energy Survey.
The Collaborating Institutions are Argonne National Laboratory, the University of California at Santa Cruz, the University of Cambridge, Centro de Investigaciones En\'ergeticas, Medioambientales y Tecnol\'ogicas–Madrid, the University of Chicago, University College London, the DES-Brazil Consortium, the University of Edinburgh, the Eidgen\"{o}ssische Technische Hochschule (ETH) Z\"{u}rich, Fermi National Accelerator Laboratory, the University of Illinois at Urbana-Champaign, the Institut de Ci\`{e}ncies de l’Espai (IEEC/CSIC), the Institut de F\'{\i}sica d’Altes Energies, Lawrence Berkeley National Laboratory, the Ludwig-Maximilians Universit\"{a}t M\"{u}nchen and the associated Excellence Cluster Universe, the University of Michigan, NSF’s NOIRLab, the University of Nottingham, the Ohio State University, the OzDES Membership Consortium, the University of Pennsylvania, the University of Portsmouth, SLAC National Accelerator Laboratory, Stanford University, the University of Sussex, and Texas A\&M University.
\end{acknowledgements}


\begin{thebibliography}{49}
\expandafter\ifx\csname natexlab\endcsname\relax\def\natexlab#1{#1}\fi

\bibitem[{{Arenou} {et~al.}(2018){Arenou}, {Luri}, {Babusiaux}, {Fabricius},
  {Helmi}, {Muraveva}, {Robin}, {Spoto}, {Vallenari}, {Antoja},
  {Cantat-Gaudin}, {Jordi}, {Leclerc}, {Reyl{\'e}}, {Romero-G{\'o}mez}, {Shih},
  {Soria}, {Barache}, {Bossini}, {Bragaglia}, {Breddels}, {Fabrizio},
  {Lambert}, {Marrese}, {Massari}, {Moitinho}, {Robichon}, {Ruiz-Dern},
  {Sordo}, {Veljanoski}, {Eyer}, {Jasniewicz}, {Pancino}, {Soubiran}, {Spagna},
  {Tanga}, {Turon}, \& {Zurbach}}]{arenouetal2018}
{Arenou}, F., {Luri}, X., {Babusiaux}, C., {et~al.} 2018, \aap, 616, A17

\bibitem[{{Balbinot} \& {Gieles}(2018)}]{bg2018}
{Balbinot}, E. \& {Gieles}, M. 2018, \mnras, 474, 2479

\bibitem[{{Baumgardt} {et~al.}(2019){Baumgardt}, {Hilker}, {Sollima}, \&
  {Bellini}}]{baumgardtetal2019}
{Baumgardt}, H., {Hilker}, M., {Sollima}, A., \& {Bellini}, A. 2019, \mnras,
  482, 5138

\bibitem[{{Bellazzini} {et~al.}(2003){Bellazzini}, {Ferraro}, \&
  {Ibata}}]{bellazzinietal2003}
{Bellazzini}, M., {Ferraro}, F.~R., \& {Ibata}, R. 2003, \aj, 125, 188

\bibitem[{{Belokurov} {et~al.}(2018){Belokurov}, {Erkal}, {Evans}, {Koposov},
  \& {Deason}}]{belokurovetal2018}
{Belokurov}, V., {Erkal}, D., {Evans}, N.~W., {Koposov}, S.~E., \& {Deason},
  A.~J. 2018, \mnras, 478, 611

\bibitem[{{Belokurov} {et~al.}(2006){Belokurov}, {Zucker}, {Evans}, {Gilmore},
  {Vidrih}, {Bramich}, {Newberg}, {Wyse}, {Irwin}, {Fellhauer}, {Hewett},
  {Walton}, {Wilkinson}, {Cole}, {Yanny}, {Rockosi}, {Beers}, {Bell},
  {Brinkmann}, {Ivezi{\'c}}, \& {Lupton}}]{belokurovetal2006b}
{Belokurov}, V., {Zucker}, D.~B., {Evans}, N.~W., {et~al.} 2006, \apjl, 642,
  L137

\bibitem[{{Carballo-Bello}(2019)}]{carballobello2019}
{Carballo-Bello}, J.~A. 2019, \mnras, 486, 1667

\bibitem[{{Carballo-Bello} {et~al.}(2017){Carballo-Bello}, {Corral-Santana},
  {Mart{\'\i}nez-Delgado}, {Sollima}, {Mu{\~n}oz}, {C{\^o}t{\'e}}, {Duffau},
  {Catelan}, \& {Grebel}}]{carballobelloetal2017}
{Carballo-Bello}, J.~A., {Corral-Santana}, J.~M., {Mart{\'\i}nez-Delgado}, D.,
  {et~al.} 2017, \mnras, 467, L91

\bibitem[{{Carballo-Bello} {et~al.}(2018){Carballo-Bello},
  {Mart{\'{\i}}nez-Delgado}, {Navarrete}, {Catelan}, {Mu{\~n}oz}, {Antoja}, \&
  {Sollima}}]{carballobelloetal2018}
{Carballo-Bello}, J.~A., {Mart{\'{\i}}nez-Delgado}, D., {Navarrete}, C.,
  {et~al.} 2018, \mnras, 474, 683

\bibitem[{{Carballo-Bello} {et~al.}(2014){Carballo-Bello}, {Sollima},
  {Mart{\'\i}nez-Delgado}, {Pila-D{\'\i}ez}, {Leaman}, {Fliri}, {Mu{\~n}oz}, \&
  {Corral-Santana}}]{carballobelloetal2014}
{Carballo-Bello}, J.~A., {Sollima}, A., {Mart{\'\i}nez-Delgado}, D., {et~al.}
  2014, \mnras, 445, 2971

\bibitem[{{Dinnbier} \& {Kroupa}(2020)}]{dk2020}
{Dinnbier}, F. \& {Kroupa}, P. 2020, arXiv e-prints, arXiv:2006.14087

\bibitem[{{Flaugher} {et~al.}(2015){Flaugher}, {Diehl}, {Honscheid}, {Abbott},
  {Alvarez}, {Angstadt}, {Annis}, {Antonik}, {Ballester}, {Beaufore},
  {Bernstein}, {Bernstein}, {Bigelow}, {Bonati}, {Boprie}, {Brooks},
  {Buckley-Geer}, {Campa}, {Cardiel-Sas}, {Castander}, {Castilla}, {Cease},
  {Cela-Ruiz}, {Chappa}, {Chi}, {Cooper}, {da Costa}, {Dede}, {Derylo},
  {DePoy}, {de Vicente}, {Doel}, {Drlica-Wagner}, {Eiting}, {Elliott}, {Emes},
  {Estrada}, {Fausti Neto}, {Finley}, {Flores}, {Frieman}, {Gerdes},
  {Gladders}, {Gregory}, {Gutierrez}, {Hao}, {Holland}, {Holm}, {Huffman},
  {Jackson}, {James}, {Jonas}, {Karcher}, {Karliner}, {Kent}, {Kessler},
  {Kozlovsky}, {Kron}, {Kubik}, {Kuehn}, {Kuhlmann}, {Kuk}, {Lahav}, {Lathrop},
  {Lee}, {Levi}, {Lewis}, {Li}, {Mandrichenko}, {Marshall}, {Martinez},
  {Merritt}, {Miquel}, {Mu{\~n}oz}, {Neilsen}, {Nichol}, {Nord}, {Ogando},
  {Olsen}, {Palaio}, {Patton}, {Peoples}, {Plazas}, {Rauch}, {Reil}, {Rheault},
  {Roe}, {Rogers}, {Roodman}, {Sanchez}, {Scarpine}, {Schindler}, {Schmidt},
  {Schmitt}, {Schubnell}, {Schultz}, {Schurter}, {Scott}, {Serrano}, {Shaw},
  {Smith}, {Soares-Santos}, {Stefanik}, {Stuermer}, {Suchyta}, {Sypniewski},
  {Tarle}, {Thaler}, {Tighe}, {Tran}, {Tucker}, {Walker}, {Wang}, {Watson},
  {Weaverdyck}, {Wester}, {Woods}, {Yanny}, \& {DES
  Collaboration}}]{flaugheretal2015}
{Flaugher}, B., {Diehl}, H.~T., {Honscheid}, K., {et~al.} 2015, \aj, 150, 150

\bibitem[{{Forbes} \& {Bridges}(2010)}]{fb2010}
{Forbes}, D.~A. \& {Bridges}, T. 2010, \mnras, 404, 1203

\bibitem[{{Fukugita} {et~al.}(1996){Fukugita}, {Ichikawa}, {Gunn}, {Doi},
  {Shimasaku}, \& {Schneider}}]{fukugitaetal1996}
{Fukugita}, M., {Ichikawa}, T., {Gunn}, J.~E., {et~al.} 1996, \aj, 111, 1748

\bibitem[{{Gaia Collaboration} {et~al.}(2018){Gaia Collaboration}, {Brown},
  {Vallenari}, {Prusti}, {de Bruijne}, {Babusiaux}, {Bailer-Jones}, {Biermann},
  {Evans}, {Eyer}, \& et~al.}]{gaiaetal2018b}
{Gaia Collaboration}, {Brown}, A.~G.~A., {Vallenari}, A., {et~al.} 2018, \aap,
  616, A1

\bibitem[{{Gaia Collaboration} {et~al.}(2016){Gaia Collaboration}, {Prusti},
  {de Bruijne}, {Brown}, {Vallenari}, {Babusiaux}, {Bailer-Jones}, {Bastian},
  {Biermann}, {Evans}, \& et~al.}]{gaiaetal2016}
{Gaia Collaboration}, {Prusti}, T., {de Bruijne}, J.~H.~J., {et~al.} 2016,
  \aap, 595, A1

\bibitem[{{Harris}(1996)}]{harris1996}
{Harris}, W.~E. 1996, \aj, 112, 1487

\bibitem[{{Helmi} {et~al.}(2018){Helmi}, {Babusiaux}, {Koppelman}, {Massari},
  {Veljanoski}, \& {Brown}}]{helmietal2018}
{Helmi}, A., {Babusiaux}, C., {Koppelman}, H.~H., {et~al.} 2018, \nat, 563, 85

\bibitem[{{Ibata} {et~al.}(1994){Ibata}, {Gilmore}, \& {Irwin}}]{ibataetal1994}
{Ibata}, R.~A., {Gilmore}, G., \& {Irwin}, M.~J. 1994, \nat, 370, 194

\bibitem[{{Ibata} {et~al.}(2018){Ibata}, {Malhan}, {Martin}, \&
  {Starkenburg}}]{ibataetal2018}
{Ibata}, R.~A., {Malhan}, K., {Martin}, N.~F., \& {Starkenburg}, E. 2018, \apj,
  865, 85

\bibitem[{{Jordi} {et~al.}(2006){Jordi}, {Grebel}, \& {Ammon}}]{jetal06}
{Jordi}, K., {Grebel}, E.~K., \& {Ammon}, K. 2006, \aap, 460, 339

\bibitem[{{Koposov} {et~al.}(2012){Koposov}, {Belokurov}, {Evans}, {Gilmore},
  {Gieles}, {Irwin}, {Lewis}, {Niederste-Ostholt}, {Pe{\~n}arrubia}, {Smith},
  {Bizyaev}, {Malanushenko}, {Malanushenko}, {Schneider}, \&
  {Wyse}}]{koposovetal2012}
{Koposov}, S.~E., {Belokurov}, V., {Evans}, N.~W., {et~al.} 2012, \apj, 750, 80

\bibitem[{{Kuzma} {et~al.}(2016){Kuzma}, {Da Costa}, {Mackey}, \&
  {Roderick}}]{kuzmaetal2016}
{Kuzma}, P.~B., {Da Costa}, G.~S., {Mackey}, A.~D., \& {Roderick}, T.~A. 2016,
  \mnras, 461, 3639

\bibitem[{{Majewski} {et~al.}(2003){Majewski}, {Skrutskie}, {Weinberg}, \&
  {Ostheimer}}]{majewskietal2003}
{Majewski}, S.~R., {Skrutskie}, M.~F., {Weinberg}, M.~D., \& {Ostheimer}, J.~C.
  2003, \apj, 599, 1082

\bibitem[{{Massari} {et~al.}(2019){Massari}, {Koppelman}, \&
  {Helmi}}]{massarietal2019}
{Massari}, D., {Koppelman}, H.~H., \& {Helmi}, A. 2019, \aap, 630, L4

\bibitem[{{Mateu} {et~al.}(2018){Mateu}, {Read}, \& {Kawata}}]{mateuetal2018}
{Mateu}, C., {Read}, J.~I., \& {Kawata}, D. 2018, \mnras, 474, 4112

\bibitem[{{Montuori} {et~al.}(2007){Montuori}, {Capuzzo-Dolcetta}, {Di Matteo},
  {Lepinette}, \& {Miocchi}}]{montuorietal2007}
{Montuori}, M., {Capuzzo-Dolcetta}, R., {Di Matteo}, P., {Lepinette}, A., \&
  {Miocchi}, P. 2007, \apj, 659, 1212

\bibitem[{{Myeong} {et~al.}(2018){Myeong}, {Evans}, {Belokurov}, {Sanders}, \&
  {Koposov}}]{myeongetal2018}
{Myeong}, G.~C., {Evans}, N.~W., {Belokurov}, V., {Sanders}, J.~L., \&
  {Koposov}, S.~E. 2018, \apjl, 863, L28

\bibitem[{{Odenkirchen} {et~al.}(2003){Odenkirchen}, {Grebel}, {Dehnen}, {Rix},
  {Yanny}, {Newberg}, {Rockosi}, {Mart{\'{\i}}nez-Delgado}, {Brinkmann}, \&
  {Pier}}]{odenetal2003}
{Odenkirchen}, M., {Grebel}, E.~K., {Dehnen}, W., {et~al.} 2003, \aj, 126, 2385

\bibitem[{{Piatti}(2017{\natexlab{a}})}]{p17c}
{Piatti}, A.~E. 2017{\natexlab{a}}, \apjl, 846, L10

\bibitem[{{Piatti}(2017{\natexlab{b}})}]{p17a}
{Piatti}, A.~E. 2017{\natexlab{b}}, \apjl, 834, L14

\bibitem[{{Piatti}(2017{\natexlab{c}})}]{p17b}
{Piatti}, A.~E. 2017{\natexlab{c}}, \mnras, 465, 2748

\bibitem[{{Piatti}(2018)}]{p2018}
{Piatti}, A.~E. 2018, \mnras, 477, 2164

\bibitem[{{Piatti}(2019)}]{piatti2019}
{Piatti}, A.~E. 2019, \apj, 882, 98

\bibitem[{{Piatti}(2020)}]{piatti2020}
{Piatti}, A.~E. 2020, arXiv e-prints, arXiv:2006.04688

\bibitem[{{Piatti} \& {Bica}(2012)}]{pb12}
{Piatti}, A.~E. \& {Bica}, E. 2012, \mnras, 425, 3085

\bibitem[{{Piatti} \& {Carballo-Bello}(2019)}]{pc2019}
{Piatti}, A.~E. \& {Carballo-Bello}, J.~A. 2019, \mnras, 485, 1029

\bibitem[{{Piatti} \& {Carballo-Bello}(2020)}]{pcb2020}
{Piatti}, A.~E. \& {Carballo-Bello}, J.~A. 2020, \aap, 637, L2

\bibitem[{{Piatti} {et~al.}(2018){Piatti}, {Cole}, \& {Emptage}}]{petal2018}
{Piatti}, A.~E., {Cole}, A.~A., \& {Emptage}, B. 2018, \mnras, 473, 105

\bibitem[{{Piatti} \& {Fern{\'a}ndez-Trincado}(2020)}]{pft2020}
{Piatti}, A.~E. \& {Fern{\'a}ndez-Trincado}, J.~G. 2020, \aap, 635, A93

\bibitem[{{Piatti} {et~al.}(2019){Piatti}, {Webb}, \&
  {Carlberg}}]{piattietal2019b}
{Piatti}, A.~E., {Webb}, J.~J., \& {Carlberg}, R.~G. 2019, \mnras, 489, 4367

\bibitem[{{Schlafly} \& {Finkbeiner}(2011)}]{sf11}
{Schlafly}, E.~F. \& {Finkbeiner}, D.~P. 2011, \apj, 737, 103

\bibitem[{{Sollima}(2020)}]{sollima2020}
{Sollima}, A. 2020, \mnras, 495, 2222

\bibitem[{{Sollima} {et~al.}(2011){Sollima}, {Mart{\'{\i}}nez-Delgado},
  {Valls-Gabaud}, \& {Pe{\~n}arrubia}}]{sollimaetal2011}
{Sollima}, A., {Mart{\'{\i}}nez-Delgado}, D., {Valls-Gabaud}, D., \&
  {Pe{\~n}arrubia}, J. 2011, \apj, 726, 47

\bibitem[{{Stetson}(1987)}]{s1987}
{Stetson}, P.~B. 1987, \pasp, 99, 191

\bibitem[{{Valdes} {et~al.}(2014){Valdes}, {Gruendl}, \& {DES
  Project}}]{valdesetal2014}
{Valdes}, F., {Gruendl}, R., \& {DES Project}. 2014, in Astronomical Society of
  the Pacific Conference Series, Vol. 485, Astronomical Data Analysis Software
  and Systems XXIII, ed. N.~{Manset} \& P.~{Forshay}, 379

\bibitem[{{Vanderbeke} {et~al.}(2015){Vanderbeke}, {De Propris}, {De Rijcke},
  {Baes}, {West}, \& {Blakeslee}}]{vanderbekeetal2015}
{Vanderbeke}, J., {De Propris}, R., {De Rijcke}, S., {et~al.} 2015, \mnras,
  450, 2692

\bibitem[{{VanderPlas} {et~al.}(2012){VanderPlas}, {Connolly}, {Ivezic}, \&
  {Gray}}]{astroml}
{VanderPlas}, J., {Connolly}, A.~J., {Ivezic}, Z., \& {Gray}, A. 2012, in
  Proceedings of Conference on Intelligent Data Understanding (CIDU), 47--54

\bibitem[{{Wang} \& {Chen}(2019)}]{wch2019}
{Wang}, S. \& {Chen}, X. 2019, \apj, 877, 116

\end{thebibliography}


\end{document}